\documentclass[12pt]{article}
\usepackage{latexsym}
\usepackage{epsfig,amssymb,euscript}
\usepackage{amsmath}
\usepackage{mathrsfs}
\numberwithin{equation}{section}  
\newsavebox{\ns}
\newsavebox{\dbrane}

\def\be{\begin{equation}}
\def\ee{\end{equation}}
\def\bea{\begin{eqnarray}}
\def\eea{\end{eqnarray}}

\newcommand{\nn}{\nonumber}

\def\Dslash{\,\,{\raise.15ex\hbox{/}\mkern-12mu D}}
\def\Dbarslash{\,\,{\raise.15ex\hbox{/}\mkern-12mu {\bar D}}}
\def\delslash{\,\,{\raise.15ex\hbox{/}\mkern-9mu \partial}}
\def\delbarslash{\,\,{\raise.15ex\hbox{/}\mkern-9mu {\bar\partial}}}
\def\pslash{\,\,{\raise.15ex\hbox{/}\mkern-9mu p}}
\def\calDslash{\,\,{\raise.15ex\hbox{/}\mkern-12mu {\cal D}}}

\newcommand\R{\mathbb{R}}
\newcommand\Z{\mathbb{Z}}

\newcommand\C{\mathbb{C}}

\newcommand\diff{\mathrm{d}}
\newcommand\ex{\mathrm{e}}
\newcommand{\de}{\partial}

\begin{document}
\begin{titlepage}
\begin{center}
\today

\vskip 1.7cm
{\Large \bf   Moduli spaces of Chern-Simons} \\[3.4mm]
{\Large \bf quiver gauge theories and AdS$_4$/CFT$_3$}

\vskip 1.5cm
{Dario Martelli$^{1*}$ ~~and~~ James Sparks$^{2}$}\\
\vskip 0.8cm

1: {\em Institute for Advanced Study\\
Einstein Drive, Princeton, NJ 08540, U.S.A.}\\
\vskip 0.8cm
2: {\em Mathematical Institute, University of Oxford,\\
 24-29 St Giles', Oxford OX1 3LB, U.K.}\\

\vskip 2.4cm

\end{center}

\vskip 1cm

\begin{abstract}
\noindent
We analyse the classical moduli spaces of supersymmetric vacua of 3d ${\cal N}=2$ Chern-Simons
quiver gauge theories. We show quite generally that the moduli space of the 3d theory
always contains a baryonic branch of a parent 4d ${\cal N}=1$ quiver gauge theory, where the
4d baryonic branch is determined by the vector of 3d Chern-Simons levels.
In particular, starting with a 4d quiver theory dual to a 3-fold singularity, for certain
general choices of
Chern-Simons levels this branch of the moduli space
of the corresponding 3d theory  is a 4-fold singularity.
Our results lead to a simple general method, using existing 4d techniques,
for constructing candidate 3d ${\cal N} = 2$ superconformal Chern-Simons quivers
 with AdS$_4$ gravity duals. As simple, but non-trivial, examples, we identify a family
of Chern-Simons quiver gauge theories which are candidate AdS$_4$/CFT$_3$ duals to an
infinite class of toric Sasaki-Einstein seven-manifolds with explicit metrics.
\end{abstract}

\vfill
\hrule width 5cm
\vskip 5mm

{\noindent $^*$ {\small On leave from: \emph{Blackett Laboratory,
Imperial College, London SW7 2AZ, U.K.}}}

\end{titlepage}
\pagestyle{plain}
\setcounter{page}{1}
\newcounter{bean}
\baselineskip18pt


\section{Introduction}

Three-dimensional Chern-Simons (CS) gauge theories coupled to matter, with ${\cal N}=2$ supersymmetry or higher, have
recently attracted considerable attention,  as prominent candidates for
field theory duals of AdS$_4$ vacua of
string and M-theory \cite{Schwarz:2004yj}. The simplest examples of these vacua are Freund-Rubin AdS$_4\times Y_7$
solutions of eleven-dimensional supergravity, where $Y_7$ is a
Sasaki-Einstein seven-manifold (or orbifold).
Such backgrounds are expected to be AdS/CFT dual to the field theory on
a large number of M2-branes at a
Calabi-Yau 4-fold
singularity. One would then like to answer the question:
what are the field theory duals
of such solutions? Of course this hinges on the open problem of what are
the degrees of freedom on the M2-branes.
Progress in this direction has been made in the recent work of \cite{ABJM} (ABJM).
The authors of the latter reference have identified
the gauge theory duals of a class of AdS$_4\times S^7/\Z_k$ backgrounds,
 showing that these are ${\cal N}=6$ (or ${\cal N}=8$)
Chern-Simons quivers with two nodes and Chern-Simons levels $(k,-k)$.
In fact, the quiver itself is precisely the
same as the 4d $\mathcal{N}=1$ model of \cite{KW}.

The corresponding situation in type IIB string theory is understood
rather better. Here one can construct large classes of ${\cal N}=1$ AdS$_5$/CFT$_4$
duals by considering $N$ D3-branes
placed at a conical Calabi-Yau 3-fold singularity $X$. In many cases the
gauge theory may be constructed from the open string
degrees of freedom living on the (fractional) branes.
In these examples  the dual theory is described by a 4d $\mathcal{N}=1$ quiver gauge theory.
 The moduli space of vacua of these theories contains
a branch (the mesonic branch) which is a symmetric product of the
Calabi-Yau singularity  $X$ one started with. The gravity dual is then expected to be
 AdS$_5\times Y_5$, where $Y_5$ is the
Sasaki-Einstein base of the Calabi-Yau  cone $X=C(Y_5)$, thus closing the circle.
The key difference with the M-theory set-up described in the paragraph above is that
D-branes in string theory are currently understood in much greater detail than M-branes in M-theory.

In this paper we analyse the classical vacuum moduli spaces (VMS) of ${\cal N}=2$ Chern-Simons quiver gauge theories with
arbitrary CS levels. These spaces in general may be rather complicated, containing several
 branches (\emph{i.e.} Coulomb, Higgs, or mixed branches). However, motivated by the situation in 4d
 and the CS quiver theory of \cite{ABJM},
 we will focus our attention on a particular branch of these theories. If the CS quiver
theories we discuss indeed have an interpretation in terms of M2-branes at a
CY 4-fold singularity, we believe it is this branch that should reproduce
the CY 4-fold as the moduli space of the transverse M2-branes.
For simplicity we will take all ranks of
the gauge groups equal to $N$
and denote this by $U(N)_{k_1}\times \cdots \times U(N)_{k_n}$,
although the results we describe may be easily generalised to the case of arbitrary ranks.
We begin with the Abelian theory $N=1$.
We show that the VMS contains a branch that is
closely related to the moduli space of a parent 4d ${\cal N}=1$
quiver theory, in a sense that we shall explain more precisely during the course of the paper.
In particular, when this parent quiver theory arises from a 3-fold singularity, for certain
general choices of
Chern-Simons levels the corresponding 3d theory
 has a branch of the  moduli space which is a 4-fold singularity.
The discussion is extended to the non-Abelian theories with little modification.

Note that, \emph{a priori}, it is not clear what are the conditions that a Chern-Simons quiver should satisfy in order
to flow to a superconformal fixed point in the infra-red (IR).
The situation ought to be more subtle than is the case in four dimensions,
where anomalies, NSVZ beta-functions,  and a-maximisation  \cite{Intriligator:2003jj}
provide important constraints on the IR dynamics.

The results of this paper are a first key step towards  identifying candidate
${\cal N}=2$  conformal Chern-Simons quiver gauge theories with AdS$_4\times Y_7$
gravity duals. In particular, they suggest a general method for constructing 3d Chern-Simons
quiver gauge theories arising from M2-branes at a given Calabi-Yau 4-fold singularity.
As an application, we discuss a family of Chern-Simons quiver gauge theories
that are candidate duals to an infinite family of explicit Sasaki-Einstein
seven-manifolds, constructed in \cite{Gauntlett:2004hh} and further analysed in
\cite{notes}.

The plan of the rest of the paper is as follows. In section \ref{lagrangians} we recall the field
content and Lagrangian of ${\cal N}=2$ Chern-Simons theories,
with product gauge group and bifundamental matter, \emph{i.e.} Chern-Simons quiver gauge theories. In section \ref{abelian}
 we analyse the VMS
of Abelian quivers. Section \ref{nonabelian} describes the extension to non-Abelian gauge groups. In section \ref{discussion}
we discuss the relevance
of our results for the construction of superconformal Chern-Simons quivers with AdS$_4$ duals.
Section \ref{examples} presents an infinite family of Chern-Simons quiver gauge
theories which are candidate AdS$_4$/CFT$_3$ duals to a corresponding family of explicit Sasaki-Einstein
seven-manifolds.

\section{Field content and Lagrangians}
\label{lagrangians}

We largely follow the notation and discussion in \cite{Schwarz:2004yj,ABJM,Gaiotto:2007qi}.
A 3d ${\cal N}=2$ vector multiplet $V$ consists of a gauge field $A_\mu$,
a scalar field $\sigma$, a two-component Dirac spinor $\chi$, and another scalar field $D$,
all transforming in the adjoint representation of the gauge group $G$.
This is simply the dimensional reduction of the usual 4d ${\cal N}=1$ vector multiplet. In particular, $\sigma$ arises from the zero mode of the
component of the vector field in the direction along which we reduce.
The matter fields $\Phi_a$ are chiral multiplets, consisting of a complex scalar $\phi_a$, a fermion $\psi_a$ and
an auxiliary scalar $F_a$, which may be in arbitrary representations $\mathcal{R}_a$ of $G$.  An ${\cal N}=2$ Lagrangian
then consists of the three terms
\bea
S \, =\, S_{\mathrm{CS}} + S_{\mathrm{matter}} + S_{\mathrm{potential}}~.
\eea
We describe each of these in turn.

We will be interested in product gauge groups of the form
\bea
\label{gaugegroup}
G \,=\, \prod_{i=1}^n U(N_i)~.
\eea
It will turn out to be important to allow different Chern-Simons levels
$k_i$ for each factor $U(N_i)$. If $V_i$ denotes the projection
of $V$ onto the $i$th gauge group factor, then in component notation the Chern-Simons Lagrangian, in Wess-Zumino gauge, takes the form
\bea
S_{\mathrm{CS}}\, = \, \sum_{i=1}^n \frac{k_i}{4\pi}\int  \mathrm{Tr} \,\left( A_i \wedge \diff A_i + \frac{2}{3} A_i\wedge A_i\wedge A_i - \bar\chi_i \chi_i +
2D_i\sigma_i \right)~.
\label{CSaction}
\eea
The Chern-Simons levels $k_i$ are quantised. In particular, for
$U(N_i)$ or $SU(N_i)$ gauge group $k_i\in\Z$ are integers if the trace in (\ref{CSaction})
is normalised in the fundamental representation.

The matter (kinetic) term takes a simple form in superspace, namely
\bea
S_{\mathrm{matter}} & = & \int \diff^3 x \diff^4\theta \sum_a \,  \mathrm{Tr} \,\bar \Phi_a \ex^{V} \Phi_a\nonumber\\
&=& \int \diff^3 x \sum_a \mathscr{D}_\mu \bar\phi_a \mathscr{D}^\mu \phi_a - \bar\phi_a\sigma^2 \phi_a + \bar\phi_a D \phi_a + \mathrm{fermions}~, \label{mataction}
\eea
where in the second line we have expanded into component fields, and we have
not written the terms involving the fermions $\psi_a$. The auxiliary
fields $\sigma$ and $D$ are here understood to act on $\phi_a$
in the appropriate representation $\mathcal{R}_a$, just as for the covariant
derivatives $\mathscr{D}_\mu$ which contain the gauge field $A_\mu$.

The superpotential term is
\bea
S_{\mathrm{potential}} & = &  \int \diff^3 x \diff^2\theta  \, W(\Phi) + c.c.\nonumber\\
&=& -\int \diff^3 x \sum_a \, \left|\frac{\partial W}{\partial \phi_a}\right|^2 + \mathrm{fermions}~.
\label{potaction}
\eea
At this stage we take the superpotential
to be an arbitrary  polynomial in the scalar fields $\phi_a$,
and we have included the couplings in the definition of $W$.
Notice that the coupling constants
 are in general not related to the Chern-Simons levels,
as is necessarily the case for ${\cal N}=3$ supersymmetry \cite{Gaiotto:2007qi}.
In particular, they may be renormalised in the IR.

The resemblance of these theories to $4d$ ${\cal N}=1$ theories should be apparent.
 Notice, however, that there are no kinetic terms for the gauge fields, which  are replaced by the CS terms.
The fields in the vector multiplets are therefore auxiliary fields.

\section{Abelian Chern-Simons quivers}
\label{abelian}

Recall that a quiver is a directed graph on $n$ nodes,
with arrow set $\mathcal{A}$ and head and tail maps
$h,t:\mathcal{A}\rightarrow
\{1,2,\ldots,n\}$. In general we associate
a gauge group factor $U(N_i)$ to node $i\in\{1,\ldots,n\}$,
with the chiral field $\Phi_a$ transforming in the
fundamental representation of the gauge group at
node $h(a)$ and the anti-fundamental representation
of the gauge group at node $t(a)$. The gauge group is
thus given by (\ref{gaugegroup}). The superpotential
$W$ is constructed as the trace of a sum of closed
oriented paths in the quiver.  The coefficients in this
sum are the (classical) superpotential couplings.

We begin by specialising to the Abelian case with $N_i=1$ for
all $i$, so that the gauge group is simply
\bea
G \,=\, U(1)^n~.
\eea
All of the gauge fields $A_i$ are hence Abelian.
The labels $a\in \mathcal{A}$ on
the chiral fields $\Phi_a$ run over arrows in the quiver,
and $\Phi_a$ has charge $+1$ under $U(1)_{h(a)}$ and
charge $-1$ under $U(1)_{t(a)}$. Furthermore,
the auxiliary fields $\sigma$ and $D$ are then $n$-component
fields, $\sigma_i$ and $D_i$.

The potential $\mathcal{V}$ for the theory is a sum of a D-term potential
and an F-term potential (given by (\ref{potaction})), so that
\bea
\mathcal{V}\,=\,\mathcal{V}_D+\mathcal{V}_F~.
\eea
Here we have defined
\bea\label{voofoo}
\mathcal{V}_F \,=\, \sum_{a\in\mathcal{A}} \, \left|\frac{\partial W}{\partial \phi_a}\right|^2~,
\eea
whereas the D-term potential takes the form
\bea
\label{pot}
\mathcal{V}_D \,=\, -\sum_{i=1}^n \frac{k_i}{2\pi} D_i \sigma_i + \sum_{a\in\mathcal{A}} |\phi_a|^2 (\sigma_{h(a)} - \sigma_{t(a)})^2
- \sum_{a\in\mathcal{A}} |\phi_a|^2 (D_{h(a)} - D_{t(a)})~.
\eea
Here the first term comes from the CS action (\ref{CSaction}),
whereas the second and third terms come from the
matter action (\ref{mataction}).
We may rewrite the last term in (\ref{pot}) as
\bea
-\sum_{a\in\mathcal{A}} |\phi_a|^2 (D_{h(a)} - D_{t(a)}) = -\sum_{i=1}^n D_i \left[\sum_{a\mid h(a)=i} |\phi_a|^2 -
\sum_{a\mid t(a)=i} |\phi_a|^2\right] = \sum_i D_i \mathcal{D}_i
\eea
where we have defined the usual 4d $\mathcal{N}=1$ D-term as
\bea\label{4dD}
\mathcal{D}_i = -\sum_{a\mid h(a)=i} |\phi_a|^2 +
\sum_{a\mid t(a)=i} |\phi_a|^2~.
\eea
Integrating out the auxiliary fields $D_i$ then immediately gives
\bea
 \mathcal{D}_i \, =\,  \frac{k_i\sigma_i}{2\pi} ~,
 \label{dterms}
\eea
where there is no summation on the right hand side.
Notice that on summing the equalities in (\ref{dterms}) over all the nodes of the quiver, the left hand side vanishes. This follows from the
fact that nothing is charged under the overall diagonal $U(1)$.
We thus find the condition
\bea
\sum_{i=1}^n k_i\sigma_i\, =\, 0~.
\label{sumCSsigma}
\eea
Substituting (\ref{dterms}) back into the action the terms involving $D_i$ cancel, because
the potential is linear in $D_i$, leaving
only the second term in (\ref{pot}). Thus
\bea
\mathcal{V}_D \,=\, \sum_{a\in\mathcal{A}} |\phi_a|^2 (\sigma_{h(a)} - \sigma_{t(a)})^2~.
\label{vodoo}
\eea

\subsection*{Supersymmetric vacua}

In vacuum the fermions are all set to zero, with the
scalar fields taking constant VEVs. The potential $\mathcal{V}$, since it is manifestly non-negative,
then has an absolute minimum at zero.
In fact since both $\mathcal{V}_D$ (\ref{vodoo}) and $\mathcal{V}_F$ (\ref{voofoo}) are
both non-negative,
each must vanish separately in a supersymmetric vacuum.

The F-term equations are simply
\bea
\frac{\de W}{\de \phi_a}\, =\, 0~.
\eea
This defines an affine algebraic set  \bea
{\cal Z}= \{ \diff W =0 \}\subset \C^{M}~,\eea where in the Abelian case at hand $M=|\mathcal{A}|$. This is exactly as for 4d $\mathcal{N}=1$ quiver gauge theories.

We next turn to the D-term equations.
Again, since (\ref{vodoo}) is a sum of non-negative terms, the potential is minimised at zero.
One set of solutions is clearly given by
\bea
\label{equalVEVs}
\sigma_1\,=\,\sigma_2\,=\,\cdots\,=\,\sigma_n\,\equiv\, s~.
\eea
Here $s\in\R$ is arbitrary. As will become clear, this is an interesting branch of the moduli space,
since the final result when the corresponding 4d quiver theory is dual to a 3-fold
singularity will  be a 4-fold singularity.
In general there could be other branches, obtained  by instead setting certain $\phi_a=0$ and thus allowing
for more general $\sigma_i$. It is simple to write down examples of quivers
which have such branches. However, we believe that for the quivers relevant for the
AdS$_4$/CFT$_3$ correspondence, it is the above branch that  should reproduce
the CY 4-fold geometry as the moduli space of transverse M2-branes.
 In any case,  we will not consider the other branches of the VMS, if indeed there are any, in the present paper.

The conditions (\ref{dterms}) then become
\bea\label{VMSeqns}
 \mathcal{D}_i\, =\, \frac{k_is}{2\pi} ~.
\eea
Note then that (\ref{sumCSsigma}) implies
\bea\label{sumCS}
\sum_{i=1}^n k_i=0~.
\eea
This is hence a necessary condition on the Chern-Simons levels for
a Chern-Simons quiver theory to admit the above vacua with $s\neq0$.
If (\ref{sumCS}) does not hold then $s$ is identically zero
and note that we reduce
to the usual 4d space of D-term equations with zero FI parameters. This would usually
be called the Higgs branch.
Indeed, the VMS equations (\ref{VMSeqns}) may be  regarded as promoting a 4d FI parameter to a VEV. The FI parameter
is $\zeta_i=k_is/2\pi$, and thus the direction is determined by the vector of CS levels,
while the scale is determined by the VEV $s$ of the auxiliary scalars. Thus, provided the vector $k=(k_1,\ldots,k_n)\neq 0$ and (\ref{sumCS}) holds, the 3d space of absolute minima
of the potential is always one real dimension higher than the 4d space of minima
for the corresponding quiver theory.

We may conveniently rewrite the 3d D-term equations (\ref{VMSeqns})
in a 4d language as follows. We begin by noting that the $n$-vector $k$ is, more invariantly,
an element of the
dual Lie algebra $\mathtt{t}_n^*\cong\R^n$ of $G=U(1)^n$, so
\bea
k\in \mathtt{t}_n^*~.
\eea
There is hence a kernel
\bea
\ker(k)\subset \mathtt{t}_n\cong \R^n~,
\eea
given by vectors that contracted with $k$ give zero. Provided $k\neq 0$, this defines a vector subspace of dimension $n-1$.
Then the 3d D-term equations (\ref{VMSeqns}) may be written as
\bea\label{bob}
\sum_{i=1}^n v_i \mathcal{D}_i=0~, \qquad v\in \ker(k)~.
\eea
Note that this gives the correct VMS even when $k=0$. Also notice that the vector
$v=(1,1,\ldots,1)\in \ker(k)$ if (\ref{sumCS}) holds. Since the D-term for this vector, and only for this direction, is identically zero, we see that (\ref{bob}) imposes $(n-2)$ linearly independent equations for $k\neq0$
satisfying (\ref{sumCS}). In fact from now on we assume the latter
conditions to hold.

\subsection*{Gauge symmetries}

In vacuum the gauge fields are set to zero\footnote{We will modify this
statement slightly below.}. Constant gauge transformations therefore map vacua to vacua, and
to compute the space of gauge-equivalent solutions we must also identify by these gauge transformations.
We have already noted that the overall diagonal $U(1)$ acts trivially,
and thus naively it seems one should quotient the space of
F-term and D-term solutions by the action of $U(1)^{n-1}\cong U(1)^n/U(1)$ to obtain the VMS.
However, there is an immediate
problem with this: the VMS would then be odd-dimensional,
which is incompatible with supersymmetry. The resolution of this apparent
puzzle becomes clear on examining the CS action more carefully, precisely
as in \cite{ABJM} (see also \cite{Imamura:2008nn}).

We define
\bea
a \,=\, \sum_{i=1}^n A_i~, \qquad\qquad   b \,=\, \frac{1}{h}\sum_{i=1}^n k_i A_i~,
\eea
where we have introduced
\bea h \,=\, \mathrm{gcd}\{k_i\}~.
\eea
The Abelian CS action for the gauge fields $A=(A_1,\ldots,A_n)$ is
\bea
S_{\mathrm{SC}}(A) \,=\, \frac{1}{4\pi}\sum_{i=1}^{n}\int k_iA_i\wedge \diff A_i~.
\eea
Now consider making the simultaneous variations
\bea
A_i\rightarrow A_i+\lambda~, \qquad i=1,\ldots,n
\eea
with $\lambda$ an arbitrary one-form. This induces the variations
\bea
\delta_\lambda a &=&  n \lambda\\
\delta_\lambda b &=& 0
\eea
where the second equation follows from (\ref{sumCS}). The variation of the CS action is hence
\bea
\delta_\lambda S_{\mathrm{CS}}(A) = \frac{2}{4\pi} \sum_{i=1}^n \int \lambda\wedge k_i\diff A_i
\eea
where note there are two terms to vary in each summand, but they give equal contributions
after integrating by parts. We may rewrite this as
\bea
\delta_\lambda S_{\mathrm{CS}}(A) \,=\, \frac{2h}{4\pi}  \int \lambda\wedge \diff b~.
\eea
We thus conclude that
\bea\label{BF}
S_{\mathrm{CS}}(A) \, = \,\frac{h}{2\pi n} \int b\wedge f + S'~,
\eea
where we have defined $f=\diff a$, and by definition
\bea
\delta_\lambda S'\,=\, 0~.
\eea
Since the overall $U(1)$ decouples from the matter, we see that the first ``$bf$''
term in the action (\ref{BF}) describes completely the
action for the gauge field $a$. We may thus introduce a Lagrange multiplier
\bea\label{Lagrange}
S_{\tau} \,=\,  -\frac{1}{2\pi}\int \diff \tau\wedge f
\eea
and treat $f$, rather than $a$, as the basic variable. Integrating out $f$ then imposes\footnote{We note
a factor of 2 difference with the corresponding equation in \cite{ABJM}.}
\bea
b \, =\, \frac{n}{h}\diff \tau~.
\eea
The gauge invariance of the theory is now
\bea\label{gaugeB}
 b\rightarrow b + \diff \theta~, \quad &&\quad \tau \rightarrow \tau + \frac{h}{n}\theta~,\\
\label{gaugeAi}
A_i\rightarrow A_i + \diff\theta_i~, \quad && \quad \sum_{i=1}^n k_i\theta_i\,=\, 0~.
\eea
The gauge transformations (\ref{gaugeAi}) are precisely those that do not act on $b$.
The transformation (\ref{gaugeB}) of $b$ instead arises from
\bea\label{Btrans}
A_i\rightarrow A_i + \diff \theta_i~, \qquad \sum_{i=1}^n k_i\theta_i = h\theta~.
\eea
Consider now the character
\bea
\label{character}
\chi_k&:& U(1)^n\rightarrow U(1)\nn\\
&;& \left(\ex^{i\theta_1},\ldots,\ex^{i\theta_n}\right)\mapsto \exp\left(i\sum_{i=1}^nk_i\theta_i\right)~.
\eea
The gauge transformation of $b$ in (\ref{gaugeB}) thus maps to
$\exp(ih\theta)$. This lies in the kernel of (\ref{character}) if and only if
\bea
\label{hquotient}
\theta \,= \, \frac{2\pi l}{h}
\eea
where $l=1,\ldots,h$. On the other hand, if we assume for the moment that $\tau$ has period $2\pi/n$, then gauge fixing
$\tau=0$ leaves a residual gauge symmetry in (\ref{gaugeB}) that is precisely the same
as (\ref{hquotient}).
The transformations (\ref{gaugeAi}) also lie in the kernel
of (\ref{character}) of course. Thus, assuming that $\tau$ has period
$2\pi/n$, we  see that the group of constant
gauge transformations acting on the VMS is precisely the kernel of (\ref{character}).
This defines an Abelian group $\ker \chi_k\subset U(1)^n$ of rank $n-1$. Note that due to (\ref{sumCS}) this
contains the overall diagonal
$U(1)$, which acts trivially. Thus the effectively acting
group of gauge symmetries is the quotient
\bea\label{Hgroup}
H_k = \ker \chi_k/U(1) \cong U(1)^{n-2}\times \Z_h~.
\eea

It thus remains to justify\footnote{We note that in \cite{Imamura:2008nn} the authors
stated explicitly that they did not have a field theory explanation for this period in their
orbifold models.} that the period of $\tau$ is indeed $2\pi/n$.
As is well-known, the periodicity for $\tau$ is related to the flux quantisation
condition on $f$ via the coupling (\ref{Lagrange}). In the above vacua
we have set all gauge fields to zero, and thus $f=0$. However, since
nothing is charged under the overall diagonal $U(1)$ gauge group, one may in fact turn on
a diagonal gauge field in the above vacua. 
To see this, note that with non-zero gauge fields but constant $\phi_a$ there is an  
additional term in the expression for energy 
\bea
\sum_{a\in \mathcal{A}} |\phi_a|^2 (A_{h(a)}-A_{t(a)})^2~.
\eea
This comes directly from the kinetic term for the $\phi_a$. Thus,
in Euclidean signature, and on  the branch we consider, 
the total energy of the vacuum vanishes if and only 
if $A_1=\cdots = A_n$, which is a diagonal flux\footnote{Equivalently, this is implied by
the equations of motion for the $\phi_a$.}. Note this is closely
related to (\ref{equalVEVs}). The quantisation condition on each $F_i$ is the usual Dirac condition
\bea
\frac{1}{2\pi}\int_\Sigma F_i \in \Z
\eea
where $\Sigma$ is any two-cycle. If $\Sigma$ is a two-sphere in $\R^3$, such a flux
would signify the presence of magnetic monopoles inside this two-sphere.
Since all $F_i$ are equal, we thus see that
\bea
\frac{1}{2\pi}\int_{\Sigma} f \in n\Z~,
\eea
which then leads to a period of $2\pi/n$ for $\tau$. Note that this analysis depends
on the branch of the vacuum moduli space we are considering. On different branches, the periodicity 
of $\tau$ may \emph{a priori} be different.

The  3d VMS, or at least the branch satisfying (\ref{equalVEVs}),
 is then the K\"ahler quotient of the space of F-term solutions $\mathcal{Z}$
by $H_k$ at moment map level zero:
\bea\label{3dVMS}
\mathscr{M}_{\mathrm{3d}}(k) \, = \, \mathcal{Z}\, //\, H_k~.
\eea
Notice this moduli space is acted on by $U(1)\cong U(1)^{n-1}/H_k$, and that a further K\"ahler quotient by this $U(1)$ would produce the usual
mesonic moduli space of the corresponding 4d theory
\bea\label{3to4}
\mathscr{M}_{\mathrm{4d}} \, =\, \mathscr{M}_{\mathrm{3d}}(k)// U(1)~.
\eea
Indeed, if one introduces an FI parameter
$\zeta\in\R$ for this $U(1)$ quotient, via (\ref{3to4}) one obtains  a family of mesonic moduli spaces
$\mathscr{M}_{\mathrm{4d}}(\zeta k)$,
labelled by $\zeta$. As reviewed for example in \cite{Martelli:2008cm},
in general the space of FI parameters for a K\"ahler quotient is a fan,
which is a set of convex polyhedral cones glued together along their boundary faces.
Inside each cone the quotient spaces are isomorphic as complex manifolds,
but have an induced K\"ahler class that depends linearly on $\zeta$.
As one moves from one cone to another along a boundary wall, the
moduli space undergoes a form of small birational transformation called a flip.
In the case at hand, the CS vector $k$ picks a particular real line through the origin in the
space of FI parameters of the corresponding 4d $\mathcal{N}=1$ theory, where
we may interpret $\zeta= s$. Thus the mesonic spaces for $\zeta>0$ are all
isomorphic, with a K\"ahler class depending linearly on $\zeta$. This will be a
(partial) resolution of the mesonic moduli space with $\zeta=0$. As one passes
to $\zeta<0$ the moduli space undergoes a flip, with again the moduli spaces
for $\zeta<0$ being all isomorphic and the K\"ahler class depending linearly on $\zeta$.
Thus the 3d VMS (\ref{3dVMS}) may be obtained by gluing this one-parameter
family of 4d mesonic moduli spaces together, with the $U(1)\cong U(1)^{n-1}/H_k$
fibred over each mesonic space in the family.

We also note that (\ref{3dVMS}) may be viewed as a (GIT) quotient of $\mathcal{Z}$ by the complexified
gauge group
\bea
H_k^\C = (\C^*)^{n-2}\times \Z_h~.
\eea
In fact we may define $\mathscr{M}_{\mathrm{3d}}(k)$ as an affine variety via
\bea\label{GIT}
\mathscr{M}_{\mathrm{3d}}(k) \, = \, \mathcal{Z}\, //\, H_k^\C \, \equiv \, \mathrm{Spec}\, \C[\mathcal{Z}]^{H_k^\C}~.
\eea
The equivalence between the two descriptions is standard -- see, for example, \cite{richard}.
Moduli spaces of quivers with relations were first introduced in
\cite{king}. Given a quiver with relations, the moduli spaces in \cite{king}
are defined by first picking a character of the gauge group, precisely as in (\ref{character}),
and then defining the holomorphic (GIT) quotient, with respect to this character
$k\in\Z^n$, of the set $\mathcal{Z}$
satisfying the relations.
This is very closely related\footnote{The moduli spaces in \cite{king} are projective
versions of (\ref{GIT}).}
to the moduli space (\ref{GIT}).

\subsubsection*{Example: the ABJM theory}

It is straightforward to recover the results of \cite{ABJM} from the above discussion.
The quiver has $n=2$ nodes with four bifundamental fields, which are grouped into two
pairs in conjugate representations of the gauge group $G=U(1)^2$. The vector of CS levels
is $(k,-k)$, in the notation of \cite{ABJM}, so $h=k$. In this Abelian
case the superpotential is identically zero, and thus the space of
F-term solutions is $\mathcal{Z}\cong \C^4$. Moreover, the group (\ref{Hgroup}) is
simply $H_k\cong \Z_{k}$, and one obtains $\mathscr{M}_{\mathrm{3d}}(k) = \C^4/\Z_{k}$ as the 3d VMS. Note in
this example that there are certainly no other branches to the VMS.
A further quotient of this space\footnote{Note the result of this further quotient does not depend on $k$.}
by the relative $U(1)$ gives the conifold singularity, which is of course the mesonic
moduli space of the 4d theory \cite{KW}.

\section{Non-Abelian Chern-Simons quivers}
\label{nonabelian}

We now return to the general case where
\bea
G=\prod_{i=1}^n U(N_i)~.
\eea
In this case $\phi_a$ is an $N_{h(a)}\times N_{t(a)}$ matrix, and $\sigma_i$ and
$D_i$ are both $N_i\times N_i$ Hermitian matrices. We denote the gauge indices
by $\alpha$, $\beta$, so that for example the matrix elements of $D_i$ are
denoted $D_{i\alpha\beta}$. Here $\alpha$,$\beta=1,\ldots,N_i$, so the
range of the gauge indices is understood to depend on $i$ in this notation.
Thus
\bea
(D \phi_a)_{\alpha\beta}\, =\, \sum_{\gamma=1}^{h(a)}D_{h(a)\alpha\gamma}\phi_{a\gamma\beta} - \sum_{\delta=1}^{t(a)}D_{t(a)\delta\beta}
\phi_{a\alpha\delta}~,
\eea
where $\alpha=1,\ldots,h(a)$, $\beta=1,\ldots,t(a)$. Note carefully the index structure.

Taking the variation of the scalar potential with respect to $D_{i\alpha\beta}$ thus gives the usual 4d D-term equation
\bea\label{naDeqn}
\frac{k_i \sigma_i}{2\pi} \, = \, -\sum_{a\mid h(a)=i}\phi_a\phi^\dagger_a+\sum_{a\mid t(a)=i}\phi^\dagger_a\phi_a\,
\equiv \, \mathcal{D}_i\eea
with $k_i\sigma_i$ playing the role of a moment map level.
Note there is no sum on $i$ here. Also note that $\sigma_i$ in (\ref{naDeqn}) is indeed
Hermitian. Substituting back into the potential, the terms involving
$D_i$ again cancel because the potential is linear in $D_i$. Since
$\sigma_i$ is Hermitian, the potential may be written
\bea
\mathcal{V}_D = \sum_{a\in\mathcal{A}} \sum_{\alpha=1}^{h(a)}\sum_{\beta=1}^{t(a)}
\left(M^\dagger_a\right)_{\beta\alpha}\left(M_{a}\right)_{\alpha\beta} = \sum_{a\in\mathcal{A}} \sum_{\alpha=1}^{h(a)}\sum_{\beta=1}^{t(a)} |M_{a\alpha\beta}|^2~.
\eea
Here we have defined
\bea
M_a \,=\, \sigma\phi_a~,
\eea
which  in matrix notation is
\bea
M_a \, = \, \sigma_{h(a)}\phi_a - \phi_a \sigma_{t(a)}~,
\eea
or in components
\bea
M_{a\alpha\beta} \,=\, \sum_{\gamma=1}^{h(a)}\sigma_{h(a)\alpha\gamma}\phi_{a\gamma\beta} - \sum_{\delta=1}^{t(a)}\sigma_{t(a)\delta\beta}
\phi_{a\alpha\delta}~.
\eea
The potential is thus minimised at
\bea\label{annoying}
M_{a}\,=\, 0~.
\eea

Recall now that the gauge group $U(N_i)$ acts on $\sigma_i$ by conjugation. So $g_i\in U(N_i)$ acts as
\bea
\sigma_i\mapsto g_i \sigma_i g_i^{-1}~.
\eea
Since $\sigma_i$ is Hermitian, it is necessarily diagonalisable by an appropriate choice of
$g_i$. The eigenvalues of $\sigma_i$ are then of course real, and in this gauge we may write
\bea
\sigma_{i\alpha\beta} = s_{i\alpha} \delta_{\alpha\beta}
\eea
where there is no sum, and $s_{i\alpha}\in \R$ are the eigenvalues. In such a gauge choice,
which always exists, we have
\bea
M_{a\alpha\beta}\, =\, (s_{h(a)\alpha} - s_{t(a)\beta})\phi_{a\alpha\beta}~.
\eea
Again, there is no sum in this formula.

\subsection*{Branches of relative dimension 1}

There are various ways of satisfying (\ref{annoying}). One solution is to take
\bea
\label{sol1}
s_{i\alpha} \,=\, s
\eea
independently of $i$ and $\alpha$. Since the overall
diagonal $U(1)$ decouples, the sum of the \emph{traces} of the 4d D-terms
$\mathcal{D}_i$ is zero. In fact this may be seen directly in the definition (\ref{naDeqn}) on noting that
\bea
\mathrm{Tr}(\phi \phi^\dagger) \,= \, \mathrm{Tr}(\phi^\dagger\phi)
\eea
for any $M\times N$ matrix $\phi$. In summing the traces of the $\mathcal{D}_i$
the above two terms appear precisely once each for each bifundamental, with opposite sign, hence the result.
Thus the branch (\ref{sol1}) exists as a solution to (\ref{naDeqn}) for non-zero $s$ only if
\bea\label{cupoftea}
\sum_{i=1}^n k_i N_i\,=\, 0~.
\eea
In fact precisely this condition arises also in the mathematics literature \cite{king}.
Indeed, notice this branch has one dimension higher than the mesonic moduli space
for the 4d theory, precisely as in \cite{king}. Thus when $N_i=N\tilde{N_i}$
this branch, when it exists, is not obviously interpreted as the moduli
space of $N$ M2-branes.
To complete the discussion of these
branches we should also analyse the gauge symmetries. Since the solution space to the
D-terms above is one dimension higher than the mesonic moduli space,
the gauge group we divide by should be codimension one in $G$.
Indeed, notice that picking (\ref{sol1}) in fact leaves
the gauge symmetry group completely unbroken. The discussion is then
very similar to the Abelian case. We may introduce
the Abelian gauge fields
\bea
a\,=\, \sum_{i=1}^n \mathrm{Tr} A_i~, \qquad b\,=\, \frac{1}{h}\sum_{i=1}^n k_i \mathrm{Tr} A_i~.
\eea
The Chern-Simons action is
\bea
S_{\mathrm{SC}}(A) \,=\, \frac{1}{4\pi}\sum_{i=1}^{n}\int k_i \mathrm{Tr} \left(A_i\wedge \diff A_i
+ \frac{2}{3}A_i^3\right)~.
\eea
Varying
\bea
A_i\rightarrow A_i + \lambda \, 1_{N_i\times N_i}
\eea
leaves $b$ invariant if (\ref{cupoftea}) holds. The variation of the CS action is then
\bea
\delta_\lambda S_{\mathrm{CS}}(A) \, = \, \frac{h}{2\pi} \sum_{i=1}^n \int \lambda \wedge \diff b~,
\eea
precisely as in the Abelian case, and hence
\bea
S_{\mathrm{CS}}(A) \, = \, \frac{h}{2\pi \sum_{i=1}^n N_i} \int b\wedge f + S'~.
\eea
Introducing $\tau$ precisely as before, and defining $M=\sum_{i=1}^n N_i$,
the gauge invariance of the theory is
\bea
 b\rightarrow b + \diff \theta~, \quad &&\quad \tau \rightarrow \tau + \frac{h}{ M}\theta~,\\
A_i\rightarrow g_iA_ig_i^{-1}-i(\diff g_i) g_i^{-1}~, \quad && \quad \prod_{i=1}^n \left(
\det g_i\right)^{k_i}=1~.\eea
The discussion of monopoles proceeds as before, implying that $\tau$ has period
$2\pi/M$, and thus the group of constant gauge symmetries $H_k$ that we quotient by is
the kernel of the character
\bea
\chi(k) &:& \prod_{i=1}^n U(N_i)\rightarrow U(1)\nn\\
&:& (g_1,\ldots,g_n) \mapsto \prod_{i=1}^n \left(\det g_i\right)^{k_i}~.
\eea
Finally, with end up with a moduli space branch that is precisely
analogous to the quiver moduli spaces in \cite{king}.
In particular, this branch has one complex dimension higher than
the mesonic moduli space one obtains by taking a K\"ahler quotient of
the space of non-Abelian F-term solutions  by
the full gauge group $G$. This is what the terminology ``relative dimension one'' means
at the beginning of this subsection.

\subsection*{Branches of relative dimension $N$}

Suppose now for simplicity\footnote{The generalisation to
arbitrary $N_i$ should be a straightforward extension.} that $N_i=N$ for all $i$.
 Then an alternative way to satisfy (\ref{annoying}) is  to take
\bea\label{sol2i}
\phi_{a\alpha\beta}&=&0, \qquad \alpha\neq \beta\\\label{sol2ii}
s_{i\alpha} &= &s_{\alpha}, \qquad \forall i~.
\eea
This imposes that the bifundamentals $\phi_a$ are all diagonal, and that the $N$ eigenvalues
of $\sigma_i$ are independent of $i$. This leads to $N$ VEVs $s_{\alpha}$, $\alpha=1,\ldots,N$.
Indeed, note that provided the $\sigma_i$ are invertible (which at a generic
point they will be) we may write (\ref{annoying}) as
\bea
\phi_a = \sigma_{h(a)}^{-1} \phi_a \sigma_{t(a)}~.
\eea
On diagonalising each $\sigma_i$ this implies that if $\phi_{a\alpha\beta}\neq 0$ we must have
\bea
s_{h(a)\alpha} = s_{t(a)\beta}~.
\eea
Thus generic $\{\phi_a\}$ reduce us to the branch in the previous subsection, whereas
diagonal, but otherwise generic, $\phi_a$ lead to (\ref{sol2i}), (\ref{sol2ii}).
Note, however, that just as for the Abelian case we might allow for even
less constrained $\sigma$ by instead further restricting certain subsets of the $\phi_a$
to be zero. This branch structure thus in general appears rather complicated.
However, for now we focus on (\ref{sol2i}), (\ref{sol2ii}).

For generic (pairwise non-equal) eigenvalues in (\ref{sol2ii}) the subgroup
of the gauge symmetry group $G$ preserving this diagonal gauge choice for $\sigma_i$ is
\bea\label{diag}
K = \left(\prod_{i=1}^n U(1)^N\right)\times S_N \cong U(1)^{nN}\times S_N~.
\eea
Here the $S_N$ permutes the diagonal elements of all the matrices, so as to
preserve (\ref{sol2ii}). When some of the eigenvalues become equal, note
that this symmetry group becomes enhanced to a non-Abelian group.
By restricting to diagonal bifundamentals (\ref{sol2i}), the
superpotential clearly reduces to $N$ copies of the $N=1$ superpotential, and
thus the space of F-term solutions is simply $\mathcal{Z}^N$.
Similarly, the CS action for the gauge group (\ref{diag})
is $N$ copies of the Abelian $N=1$ CS action, with the overall
$U(1)$ decoupling in each copy separately. Thus
one clearly obtains $N$ copies of the $N=1$ VMS, with the
permutation group $S_N$ in (\ref{diag}) simply permuting the
copies. Thus this branch of the VMS is  the symmetric product
\bea\label{NVMS}
\mathscr{M}_{\mathrm{3d},N}(k) \, = \, \mathrm{Sym}^N \, \mathscr{M}_{\mathrm{3d},1}(k)
\eea
where $\mathscr{M}_{\mathrm{3d},1}(k)$ is the Abelian moduli space.
 Notice this branch is the moduli space found in \cite{ABJM}
for the ABJM theory.
Note also that this moduli space is $N$ complex dimensions higher than
the mesonic moduli space, compared to 1 complex dimension higher
for the branch discussed in the previous subsection.
It seems reasonable, given the discussion above, that the
various branches that generally exist inbetween these two extremes
have relative dimensions between 1 and $N$, and thus
the branch (\ref{NVMS}) is in fact the highest dimensional branch
of the full VMS. It has a natural physical interpretation as the moduli
space of $N$ point-like objects on $\mathscr{M}_{\mathrm{3d}}(k)=\mathscr{M}_{\mathrm{3d},1}(k)$.
 The full VMS appears to be quite a complicated object in general.
It would be interesting
to investigate more carefully the structure we have outlined above.
In particular, there may be a more elegant method
for analysing the full moduli space than the simple discussion above.

\section{In search of conformal Chern-Simons quivers}
\label{discussion}

The results we have discussed so far 
in this paper are rather  general: we have discussed
the classical vacuum moduli spaces of ${\cal N}=2$ CS quivers,
where the bifundamental matter and superpotential are arbitrary.
When the Chern-Simons quiver arises from a parent 4d quiver gauge theory dual
to a 3-fold singularity, namely the
matter content and interactions of the 3d theory are
formally the same as those of the 4d theory, our results imply that the VMS contains
(the symmetric product of) a complex four-dimensional branch of the corresponding
baryonic moduli space. More precicisely, we have found that a necessary condition for
such supersymmetric
vacua to exist is that the sum of the CS levels vanishes
\bea
\sum_{i=1}^n k_i \, =\, 0~.
\label{itsnew}
\eea

The space ${\cal Z}$ of F-term solutions is in general a fairly complicated object, with several branches of different dimension. For the class of 4d quiver gauge theories arising from D3-branes at toric Calabi-Yau singularities, this space has recently been studied\footnote{In \cite{Forcella:2008bb} this is referred to as the master space, and is denoted ${\cal F}^\flat$.}
in \cite{Forcella:2008bb}.
In the latter reference it is shown that in these examples, with $N=1$,
  ${\cal Z}$ is  a  complex $(n+2)$-dimensional affine toric variety. Moreover,  there exists a particular branch (the irreducible component), that  is argued to be itself an affine Calabi-Yau toric variety. This may be described as a K\"ahler quotient at level zero $^{\mathrm{irr}}{\cal Z}=\C^c//U(1)^{c-n-2}$, where $c$ is a number determined from the data of the quiver. The mesonic moduli space
of the theory is obtained by performing a further K\"ahler quotient, and results in the Calabi-Yau 3-fold
\bea
\mathscr{M}_{4d} \, = \, ^{\mathrm{irr}}{\cal Z}//U(1)^{n-1}~.
\eea
Taking the same quiver and replacing the kinetic terms for the gauge fields with Chern-Simons terms with CS level vector $k=(k_1,\dots, k_n)$ obeying (\ref{itsnew}), we obtain instead a branch of the 3d VMS, namely
\bea
\mathscr{M}_{3d}(k) \, = \,  ^{\mathrm{irr}}{\cal Z}//H_k~.
\eea
This a Calabi-Yau \emph{4-fold}. To see this, notice that the group $U(1)^{n-1}$ necessarily preserves the holomorphic volume form of $^{\mathrm{irr}}{\cal Z}$, since $\mathscr{M}_{4d}$ is Calabi-Yau. Thus, in particular, the subgroup  $H_k$ preserves this volume form also.

Ultimately, we are interested in conformal field theories. These are candidate gauge theory duals
of AdS$_4$ vacua of string or M-theory. Such CFTs, however, will
generically be strongly coupled\footnote{The ABJM theory is a notable exception, since it has a weakly coupled limit for large $k$.},
and at present there are  no techniques available to perform independent field theory calculations.
Note this is different from four dimensions, where a-maximisation \cite{Intriligator:2003jj} is an important tool for testing the existence of conjectured IR fixed points.
Using the AdS/CFT correspondence, the issue of conformal invariance in the IR may be translated
 into the question of whether the theory has an AdS$_4\times Y_7$  gravity dual, where $Y_7$ is a Sasaki-Einstein seven-manifold. These backgrounds arise as the near-horizon limit of a large number of M2-branes, placed at the singularity of the Calabi-Yau cone $C(Y_7)$. Thus, a necessary condition for this situation to hold is that the
3d gauge theory contains this Calabi-Yau 4-fold as a (generic) component of its VMS.
This suggests that (\ref{itsnew}) is in fact a necessary condition for
conformal invariance.
However, there may be additional conditions, yet to be discovered,
that a Chern-Simons quiver gauge theory should satisfy
in order to flow to a dual conformal fixed point in the IR. Understanding these conditions is clearly an interesting direction for future research.

Notice that different theories may lead to the same moduli space $\mathscr{M}_{3d}(k)$. This may sound surprising at first, but
one should bear in mind that this phenomenon already exists in 4d. There, Seiberg duality implies that
different gauge theories all flow to the same conformal field theory in the IR. In fact, instead of thinking of 
gauge theory duals of some AdS$_5$ solution, we should more precisely think of classes of 
Seiberg-dual gauge theories. Similar dualities
exist for 3d theories -- see \cite{Giveon:2008zn} for a recent discussion.
However, we are led to consider the possibility that for appropriate values of the Chern-Simons levels, 
(infinite) families of 4d quivers (\emph{e.g.} the $Y^{p,q}$ quivers \cite{Benvenuti:2004dy}), 
may all have the same AdS$_4$ duals, when viewed as 3d Chern-Simons quivers. 
It is unclear to us whether this will actually be the case, or rather further analysis
will reveal that these quivers do not flow to conformal field theories in three dimensions.
It will be interesting to analyse this further.

The results discussed here lead to a simple general method
for constructing candidate\footnote{It is only a candidate because it is possible that the 
3d theory will have to obey additional properties
in order to flow to a dual conformal field theory in the IR, 
as discussed above. Our analysis here is purely classical.}
3d ${\cal N} = 2$ superconformal Chern-Simons quivers
with AdS$_4$ gravity duals, using well-developed 4d techniques.
We illustrate this in the following section.


\section{Example: Chern-Simons quiver gauge theories for the $Y^{p,k}(\C P^2)$ metrics}
\label{examples}

In this final section we discuss a simple class of examples of the construction described in this
paper\footnote{This section has been added in a revised version (v2) of the paper.
In reference \cite{hanzaf}, which appeared before the present version but after the first
version,
the authors also discuss the quiver below. However, 
they did not make the connection with the explicit
metrics in \cite{Gauntlett:2004hh,notes}.}.
These are candidate gauge theory duals of the explicit Sasaki-Einstein metrics  
presented in
\cite{Gauntlett:2004hh}. Other examples may be treated in a similar manner  --
we briefly comment on various simple extensions at the end of the paper.

\begin{figure}[ht!]
\epsfxsize = 5cm\centerline{\epsfbox{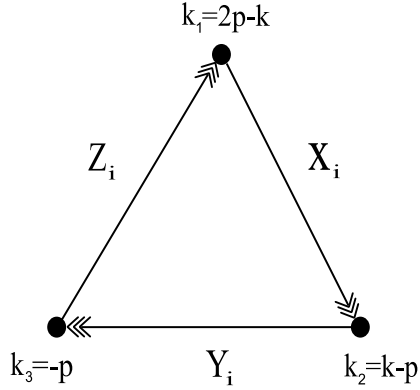}}
\caption{Quiver diagram for the candidate CS gauge theory duals of $Y^{p,k}(\C P^2)$.}\label{quiverpic}
\end{figure}

Recall from \cite{Gauntlett:2004hh,notes} that the $Y^{p,k}(\C P^2)$ metrics enjoy an $SU(3)\times U(1)^2$
isometry, and that the corresponding Calabi-Yau cones are described by a GLSM on $\C^5$, with a set of $U(1)$ 
charges characterised by two integers. 
These  properties motivate considering a quiver gauge theory with 3 nodes and  $SU(3)$ symmetry. As we shall see, 
this seemingly naive hypothesis  leads to a  consistent picture.
We thus  begin with the 4d quiver gauge theory that is AdS$_5$/CFT$_4$ dual to the orbifold $S^5/\Z_3$,
where the $\Z_3\subset U(1)$ is embedded along the Hopf $U(1)$. Equivalently, this is the theory
on $N$ D3-branes placed at the singularity of the canonical complex cone over $\mathbb{C} P^2$, which is the orbifold $\C^3/\Z_3$. The quiver has 3 nodes, with a $U(N)$ gauge group at each node,
 and 9 bifundamental fields, $X_i$, $Y_i$, $Z_i$, $i=1,2,3$,
going from nodes 1 to 2, 2 to 3, and 3 to 1, respectively. This is shown in Figure \ref{quiverpic}.

The superpotential takes the $SU(3)$-invariant form
\bea W \,=\, \epsilon_{ijk}\mathrm{Tr}\left(X_iY_jZ_k\right)~.
\eea
The F-term equations $\diff W=0$ are hence
\bea\label{Fterms}
X_iY_j \, =\, X_jY_i~,\qquad Y_iZ_j\, =\, Y_jZ_i~,\qquad Z_iX_j\,=\, Z_jX_i~.
\eea
Notice the equations with $i=j$ are redundant. Henceforth we set $N=1$, so that the bifundamental VEVs are simply coordinates on $\C^9$. Since the equations (\ref{Fterms}) set one monomial equal to another
monomial, it is a standard result that the affine variety $\mathcal{Z}=\{\diff W=0\}\subset \C^9$ is a toric variety -- see, for example, \cite{cox}.

We may equivalently realise $\mathcal{Z}$ as the affine GIT quotient (or equivalently K\"ahler quotient by $U(1)\subset \C^*$ at level zero)
\bea\label{C6}
\mathcal{Z}\, =\, \C^6\, // \, \C^*_{(1,1,1,-1,-1,-1)}~.
\eea
Here the subscript vector denotes the weights of the $\C^*$ action on $\C^6$. Thus, if we introduce coordinates $u_i,v_i$ on $\C^6$, $i=1,2,3$, then the $u_i$ have charges $+1$ and the $v_i$ have charges $-1$ under
the $\C^*$ action. The quotient (\ref{C6}) is then defined algebraically as
\bea
\mathcal{Z}\, =\, \mathrm{Spec}\, \C[u_1,u_2,u_3,v_1,v_2,v_3]^{\C^*}~.
\eea
In words, $\mathcal{Z}$ is the affine variety whose holomorphic functions are precisely the
 $\C^*$-invariant functions on $\C^6$. This ring of invariant functions is spanned by
\bea
x_i\, =\, u_1v_i~, \quad \qquad y_i\,=\,u_2v_i~,\qquad \quad z_i\,=\,u_3v_i~.
\eea
This embeds $\mathcal{Z}$ into $\C^9$, and one easily sees that the relations between
the $x_i$, $y_i$ and $z_i$ are indeed precisely the F-term relations (\ref{Fterms}).
This proves the equivalence\footnote{See also \cite{Forcella:2008bb}.} of the two descriptions of $\mathcal{Z}$.

For the 3d CS quiver theory, we introduce a CS vector $(k_1,k_2,k_3)$, where $k_3=-k_1-k_2$,
so that (\ref{sumCS}) holds.
In order to obtain the 4d VMS, which is the orbifold $\C^3/\Z_3$, we would quotient
$\mathcal{Z}$ by $(\C^*)^3/\C^*\cong (\C^*)^2$.
In 4d terms, these are the two anomalous\footnote{There is also a discrete non-anomalous
baryonic symmetry. A complete discussion of the
discrete symmetries of this theory may be found in \cite{sergei}.}
baryonic symmetries of the theory. However, to compute the moduli space of the 3d CS theory
we instead quotient by the kernel of the map
\bea\label{mappy}
(\C^*)^3 \ni (\lambda_1,\lambda_2,\lambda_3) \mapsto \lambda_1^{k_1}\lambda_2^{k_2}\lambda_3^{k_3}\in \C^*~.
\eea
This kernel, after dividing by the diagonal $\C^*$ which acts trivially, is isomorphic
to $\C^*\times \Z_h$, where $h=\mathrm{gcd}(k_1,k_2)$. For simplicity we
 begin by choosing  the CS levels so that $h=1$.
The non-trivial $\C^*$ in the kernel of (\ref{mappy}) is then generated by the weight
vector $(-k_2,k_1,0)$. The charges of the bifundamentals $X_i$, $Y_i$, $Z_i$
under a $\C^*$ action with weights $(q_1,q_2,q_3)\in\Z^3$ are $q_2-q_1$, $q_3-q_2$, $q_1-q_3$,
respectively. Thus the charges under the $\C^*$ of interest are $k_1+k_2$, $-k_1$, $-k_2$,
respectively. This determines a $\C^*$ action on $\mathcal{Z}$, which we may lift to an action
on $\C^6$ by assigning charges $(k_1+k_2,-k_1,-k_2,0,0,0)$ to the coordinates
$(u_1,u_2,u_3,v_1,v_2,v_3)$ on $\C^6$. Altogether, we thus see that the 3d VMS,
for $\mathrm{gcd}(k_1,k_2)=1$, is the affine quotient of $\C^6$ by $(\C^*)^2$ with charges
\bea
Q\, =\, \left(\begin{array}{cccccc}1 & 1 & 1 & -1 & -1 & -1\\
k_1+k_2 & -k_1 & -k_2 & 0 & 0 & 0
\end{array}\right)~.
\eea
Notice that this quotient preserves the $SU(3)$ symmetry. We now make an $SL(2,\Z)$ transformation via
\bea
\left(\begin{array}{cc}
1 & -k_1-k_2 \\
0 & 1
\end{array}\right)~,
\eea
thus giving an equivalent quotient with charges
\bea
Q'\, = \,\left(\begin{array}{cccccc}
1 & 1 & 1 & -1 & -1 & -1\\
0  & -2k_1-k_2  & -k_1-2k_2 & k_1+k_2 & k_1+k_2 & k_1+k_2
\end{array}\right)~.
\eea
We then change variables by defining
\bea
k_1 &=& 2p-k\nn\\
k_2 & =& k-p\label{defk12}
\eea
to obtain
\bea
Q'\, =\, \left(\begin{array}{cccccc}
1 & 1 & 1 & -1 & -1 & -1\\
0  & -3p +k   & -k & p & p & p\end{array}\right)~.
\eea
Thus
\bea
\mathscr{M}_{\mathrm{3d}}(2p-k,k-p,-p) \,  = \,  \C^6\, //\, (\C^*)^2_{Q'}~.
\eea

This realises the VMS $\mathscr{M}_{\mathrm{3d}}$ explicitly as a toric CY 4-fold.
We may compute the toric diagram\footnote{We refer to \cite{notes} for a review of the
relevant toric geometry.} in the usual manner, obtaining the normal vectors
\bea
&& w_0 \,=\, [0,0,k-p]~, \qquad w_1 \, =\,  [0,0,0]~, \qquad w_2 \,= \,[0,0,p]~, \nn \\
&& w_3 \,= \,[1,0,0]~, \qquad  \quad  \ \  w_4 \,= \,[0,1,0]~, \quad  \quad \; w_5 \, =\,  [-1,-1,k]~.\label{5legs}
\eea
We now note that the 5 vectors $w_1,\ldots,w_5$ precisely define the toric diagram obtained
 in \cite{notes} for the cone over the explicit Sasaki-Einstein manifolds $Y^{p,k}(\C P^2)$ of
\cite{Gauntlett:2004hh}. The vertices $w_1,\ldots,w_5$ define a compact
convex lattice polytope $\mathcal{P}$ in $\R^3$, shown in Figure 1 of reference \cite{notes}.
Of course, in (\ref{5legs}) we have 6 vectors, after including the
vertex $w_0$. However, adding this vertex will define the \emph{same} affine toric variety as $\mathcal{P}$,
provided the vertex lies inside the polytope $\mathcal{P}$. In this case,
we simply obtain a non-minimal presentation of the
toric variety, with the additional vertex $w_0$ corresponding to a blow-up mode of the singularity.
One easily sees from \cite{notes} that $w_0$ lies inside $\mathcal{P}$ provided  $p\leq k\leq 2p$. Thus,
provided $k$ lies within this range, the VMS for the CS quiver gauge theory above with
CS levels $(2p-k,k-p,-p)$ is precisely the cone over the explicit Sasaki-Einstein manifold
$Y^{p,k}(\C P^2)$.

It was shown in \cite{Gauntlett:2004hh,notes} that the metrics $Y^{p,k}(\C P^2)$
exist  for integers $p,k$ satisfying the bounds $\frac{3}{2}p \leq k \leq 3 p$. In fact,
the lower bound here is just a convention. From the explicit analysis in
\cite{notes}, one sees that the range of $k$ may be extended to lie in the interval
\bea
0 \leq k \leq 3 p ~.\label{newoldlimits}
\eea
However, notice that the GLSM quotient is manifestly invariant under the exchange of
$k$ with $3p-k$.  It is satisfying to find  that the explicit
 metrics \cite{Gauntlett:2004hh,notes} are also invariant under this exchange. This may be verified by
observing that under this transformation the roots $x_1,
x_2$ (recall $h=3$ in the notation of \cite{notes}) of the equations (2.20) in \cite{notes} are interchanged.
Thus solutions with  $k\in [0,\frac{3}{2}p]$ are equivalent to solutions with
 $k\in [\frac{3}{2}p,3p]$, which is the range considered in \cite{Gauntlett:2004hh}. Hence,
without loss of generality, we may take $k\in [\frac{3}{2}p,3p]$.

To conclude, we have thus
constructed an infinite family of CS quiver gauge theories which have explicit candidate
Sasaki-Einstein duals, for values of $p,k$ such that\footnote{Equivalently, $p \leq k \leq \frac{3}{2} p$.}
\bea
\frac{3}{2}p \leq k \leq 2 p ~.\label{newlimits}
\eea
Notice then that $k_1$ and $k_2$ are non-negative.
Given a quiver with CS levels $(k_1\geq 0, k_2 \geq 0, k_3 \leq 0)$, we may easily determine
the values of $p$, $k$ of the corresponding dual Sasaki-Einstein metric. Using (\ref{defk12}), we find $p=k_1+k_2$ and $k=k_1 + 2k_2$. Of course, we may equally pick $p=k_1+k_2$ and $k=2k_1 + k_2$.  However, from the
discussion above, the two choices are in fact completely equivalent, both for the VMS and for the explicit
metrics.

It is interesting to examine the two limiting cases of the interval (\ref{newlimits}).
At the lower bound, $p=2r$, $k=3r$, the CS levels are $(r,r,-2r)$, and the VMS is then a
$\Z_r$ orbifold of the quotient of $\C^5$ by the $\C^*$ with charges
\bea
(2,2,2,-3,-3)~.
\eea
Notice this case is symmetric under exchanging $k$ and $3p-k$. In fact, this is 
the cone over the homogeneous Sasaki-Einstein manifold $Y^{2,3}(\C P^2) = M^{3,2}$ \cite{notes}. 
The gauge
 theory we are proposing here as being dual to this manifold is then different from the proposal
made in \cite{italians}. For $k=2p$ we obtain the CS level vector $(0,p,-p)$, and
the VMS is then a $\Z_p$ orbifold of the quotient of $\C^5$ by the $\C^*$ with charges
\bea
(1,1,1,-2,-1)~.\label{oky}
\eea
Notice that $Y^{1,2}(\C P^2)$ is in a sense the first non-trivial member of the $Y^{p,k}(\C P^2)$
family of metrics. Numerical values for the volumes of this particular example were given in \cite{notes}.
It would be interesting to construct the CS quivers dual
to the metrics in \cite{Gauntlett:2004hh} with $k\in (2p,3p]$.

The only check of the conjectured duality we can make  at the time of writing is that the
VMSs of the CS quiver theories contain the corresponding Calabi-Yau 4-fold geometries as
a branch\footnote{Notice that the scalar holomorphic Kaluza-Klein spectrum will automatically be in 1-1 correspondence with the
holomorphic functions on the Calabi-Yau cone \cite{bpsmesons}, or its $N$-fold symmetric product
\cite{counting,dualg}; therefore, this matching does not constitute an independent check.}.
Combining the geometric discussion above with 
the results in \cite{notes}, it is straightforward\footnote{Note that in doing so one is ignoring the subtleties involved in constructing baryon-like operators in 3d Chern-Simons quivers, where the gauge groups of the UV theory 
are $U(N)$.} to give an assignment of R-charges of the nine bifundamental fields $X_i,Y_i,Z_i$.
It would be extremely desirable to check the proposed duality further by performing a
suitable purely field-theoretic calculation, in the spirit of $a$-maximisation.

Given the above construction, it is natural to conjecture that the CS quiver gauge theories dual
 to the $Y^{p,k}(B_4)$ manifolds constructed in \cite{Gauntlett:2004hh}, where $B_4$ may be any
  K\"ahler-Einstein 4-manifold, are described precisely by the 4d quivers for the corresponding
canonical complex cones over $B_4$. The remaining possibilities for $B_4$ are $\C P^1\times \C P^1$,
which was also discussed extensively in \cite{notes},
and the del Pezzo surfaces $dP_n$, $n=3,\ldots,8$.
Notice that the six-dimensional manifolds $M_6$,
obtained in the reduction to type IIA described in \cite{notes},
are precisely a projective version of these complex cones; these are obtained by compactifying the $\C^*$ fibers to $\C P^1$, as described in
\cite{Gauntlett:2004hh}. We leave a fuller investigation of these
models for future work.

\subsubsection*{Note added} Whilst finalising this paper for submission to the
archive we received the preprint \cite{alessandro}, which contains comments
related to the results presented here. We are grateful to the authors of \cite{alessandro}
for informing us about the completion of their work. After submitting 
the first version of this paper to the archive, the work \cite{hanzaf} appeared. 
This also has overlap with the results we have presented.

\subsection*{Acknowledgments}
\noindent
We thank Marcus Benna, Oren Bergman, Amihay Hanany, Daniel Jafferis,
Juan Maldacena, Yuji Tachikawa and
Alessandro Tomasiello for discussions. D. M.  acknowledges support from NSF grant PHY-0503584. J. F. S.
is funded by a Royal Society University Research Fellowship.

\end{document}